\documentclass[twocolumn,a4paper,amsmath,amssymb,superscriptaddress, showpacs]{revtex4}
\usepackage{graphicx}
\usepackage{color}

\begin{document}

\newcommand{\ketbra}[2]{\ensuremath{|#1\rangle\langle#2|}}
\newcommand{\ket}[1]{\ensuremath{|#1\rangle}}
\newcommand{\braket}[1]{\ensuremath{\left\langle{#1}\right\rangle}}
\newcommand{\beq}{\begin{equation}}
\newcommand{\eeq}{\end{equation}}
\newcommand{\bea}{\begin{eqnarray}}
\newcommand{\eea}{\end{eqnarray}}

\title{Dynamical creation of a supersolid in asymmetric mixtures of bosons}

\author{Tassilo Keilmann}
\affiliation{Max-Planck-Institut f\"{u}r Quantenoptik,
Hans-Kopfermann-Str. 1, Garching, D-85748, Germany}

\author{Ignacio Cirac}
\affiliation{Max-Planck-Institut f\"{u}r Quantenoptik,
Hans-Kopfermann-Str. 1, Garching, D-85748, Germany}

\author{Tommaso Roscilde}
\affiliation{Max-Planck-Institut f\"{u}r Quantenoptik,
Hans-Kopfermann-Str. 1, Garching, D-85748, Germany}
\affiliation{Ecole Normale Sup\'{e}rieure,
 46 All\'ee d'Italie, F-69007 Lyon, France}

\begin{abstract}
We propose a scheme to dynamically create a supersolid state in an optical lattice, using an attractive mixture of mass-imbalanced bosons.
Starting from a ``molecular" quantum crystal, supersolidity is induced dynamically as an out-of-equilibrium state. 
When neighboring molecular wavefunctions overlap, both bosonic species simultaneously exhibit quasi-condensation and long-range 
solid order, which is stabilized by their mass imbalance.
Supersolidity appears in a perfect one-dimensional crystal, without the requirement of doping.
Our model can be realized in present experiments with bosonic mixtures that feature simple on-site interactions, clearing the path to the observation 
of supersolidity.
\end{abstract}


\maketitle

The intriguing possibility of creating a quantum hybrid exhibiting both superflow and solidity has been envisioned long ago \cite{leggettches}. 
However, its experimental observation remains elusive. The quest for supersolidity has been strongly
 revitalized by recent experiments showing possible evidence for a non-zero superfluid fraction
 present in solid $^4$He \cite{kimbeam}. Yet, several theoretical results \cite{Prokofev07} 
appear to rule out the presence of condensation in the pure
solid phase of $^4$He, and various experiments \cite{Chan08} show indeed a strong dependence 
of the superfluid fraction on extrinsic effects, such as $^3$He impurities and dislocations.
While the experimental findings on bulk $^4$He remain controversial, optical lattice setups \cite{Blochetal08} offer the advantages of high sample purity and experimental control to directly pin down a supersolid state via standard measurement techniques.
A variety of lattice
boson models with strong finite-range interactions has been recently shown to display crystalline 
order and supersolidity upon doping
the crystal state away from commensurate filling \cite{Prokofev07,jaksch}; yet sizable interactions with a finite range 
are generally not available in current cold-atom experiments. 
Such interactions can be in principle
obtained effectively by adding a second atomic species of fermions \cite{Titvinidz}, 
which, however, does not participate in the condensate state, in a way similar to the nuclei 
forming the lattice of a superconductor without participating in the condensate of electron pairs.  

\begin{figure} [!h]
  \centering
  \includegraphics[width=0.9\linewidth]{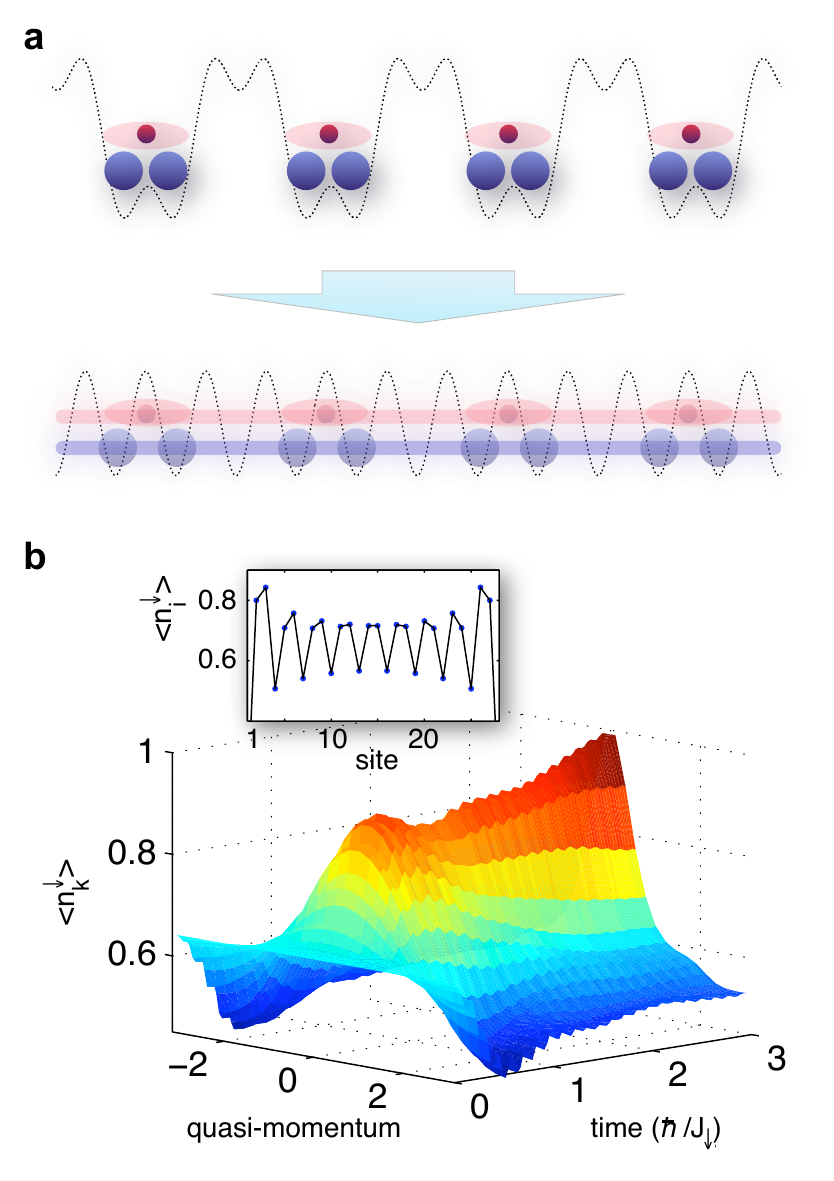}
  \vspace{-0.5cm}
  \caption{\textbf{Dynamical onset of supersolidity by quantum quenching a mixture of light and heavy bosons.}
  \textbf{(a)} A product state of bosonic trimers is the 
  initial state of the evolution (larger symbols represent the $\downarrow$-bosons); switching off one of the superlattice components
  leads to a supersolid state in which the particles delocalize into a (quasi-)condensate while maintaining the original solid pattern without imperfections.  
  \textbf{(b)} Momentum profile of the $\downarrow$-bosons, $\langle n^\downarrow_k \rangle$ vs. time in units of hopping events $\hbar / J_\downarrow$. 
  A quasi-condensate peak develops rapidly. Inset: Density distribution $\langle n^\downarrow_i \rangle$ averaged over the last third of the evolution time, showing that crystalline order is conserved in the system. (The simulation parameters are  $L=28$, $N_{\downarrow} = 18$, $N_{\uparrow} = 9$, $J_\downarrow/J_\uparrow=0.1, U/J_\uparrow=3.0$.) 
}
  \label{fig:evolution}
\end{figure}
 
 Here we demonstrate theoretically a new route to supersolidity, realized as the out-of-equilibrium
 state of a realistic lattice-boson model after a so-called  ``quantum quench" (a sudden change 
 in the Hamiltonian). The equilibrium Hamiltonian of the model before the quench realizes a ``molecular crystal" phase 
 characterized by the crystallization of atomic trimers made of two mass-imbalanced
 bosonic species. Starting from a solid of tightly-bound trimers and suddenly changing the system
 Hamiltonian, the evolution induces broadening 
 and overlap of neighboring molecular wavefunctions leading to quasi-condensation of all atomic species,
 while crystalline order is maintained
 (Fig. \ref{fig:evolution}).
 Our model requires only local on-site interactions as currently featured by neutral cold atoms,  
 which make the observation of a supersolid state a realistic and viable goal. 

 \begin{figure}
  \centering
  \includegraphics[width=0.9\linewidth]{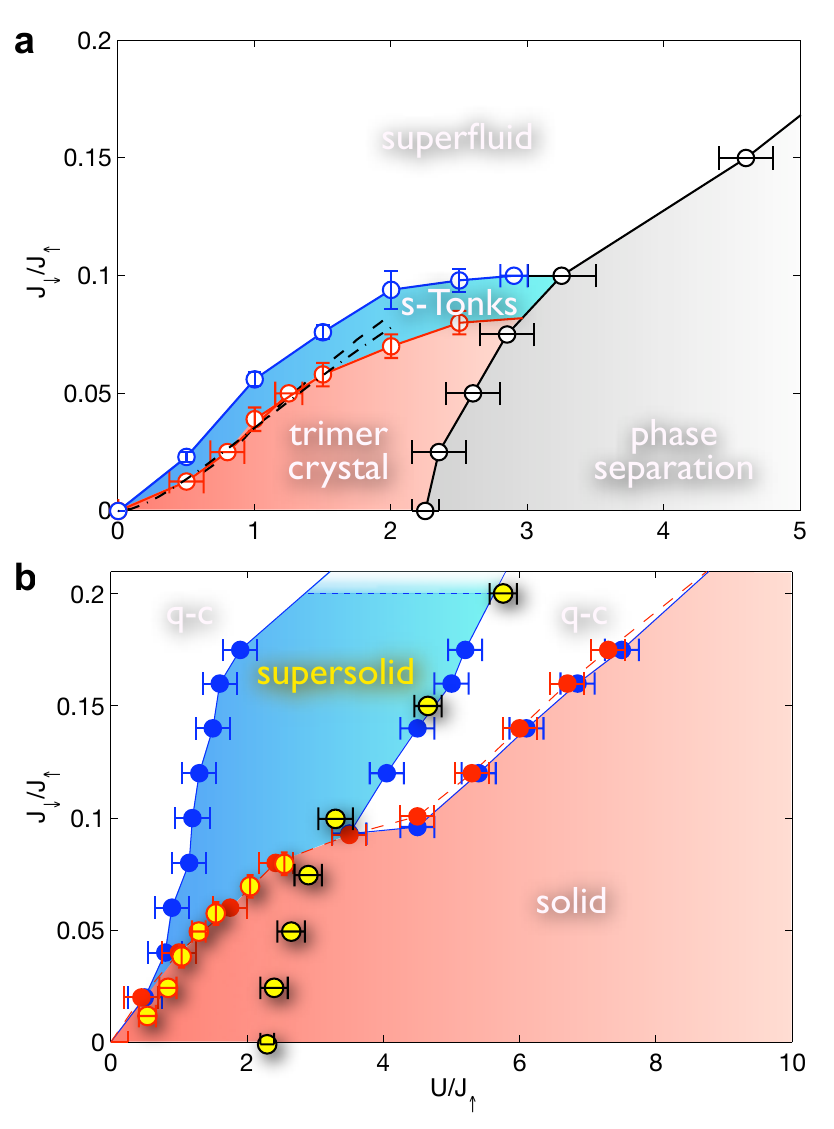}
  \vspace{-0.5cm}
  \caption{\textbf{Phase diagrams in and out of equilibrium.}
\textbf{(a)} Equilibrium phase diagram (empty circles). The dash-dotted line represents the points where the hopping of the
$\downarrow$-bosons, $J_\downarrow$, 
  overcomes the energy gap to crystal dislocations, giving rise to the solid/super-Tonks (s-Tonks) transition. 
The dashed line marks the points where a single-trimer wavefunction spreads over 2.8
 sites. 
\textbf{(b)} Out-of-equilibrium phase diagram. 
An extended supersolid phase exists in the transient state attained after the 
quantum quench.  
Blue symbols delimit the boundaries of the solid phase, red symbols mark the lower boundary for the quasi-condensed (q-c) phase. 
The overlap of both phases (blue shaded region) is identified as the supersolid phase.
The yellow-filled symbols correspond to equilibrium data points. 
The lower boundary of the superfluid/super-Tonks region of the
equilibrium phase diagram is seen to coincide with the lower boundary of the supersolid
region out of equilibrium. 
}
  \label{fig:phases}
\end{figure}

 We consider two bosonic species  ($\sigma=\uparrow,\downarrow$) tightly confined in two
transverse spatial dimensions and loaded in an optical lattice potential in the third dimension.   
In the limit of a deep optical lattice, the dynamics of the atoms can be described by a 
model of lattice hardcore bosons in one dimension \cite{jaksch98, Paredes04}
\begin{equation}
  \label{BHM}
{\cal H} = -\sum_{i,\sigma} J_{\sigma} \left( b_{i,\sigma}^\dagger b_{i+1,\sigma} + {\rm h.c.} \right)
 - U \sum_{i}  n_{i,\uparrow} n_{i,\downarrow}.
\end{equation}

 Here the operator $b^\dagger_{i\sigma}$ ($b_{i\sigma}$) creates (annihilates) a hardcore boson of species $\sigma$ on site $i$
of a chain of length $L$, 
and it obeys the on-site anticommutation relations $\{ b_{i\sigma}, b_{i\sigma}^\dagger  \} = 1$. 
$n_{i \sigma} \equiv b_{i\sigma}^\dagger b_{i\sigma}$ 
is the number operator. Throughout this Letter we restrict ourselves to the case of attractive on-site interactions $U>0$ 
and to the case of mass imbalance, $J_{\uparrow} > J_{ \downarrow}$. Moreover we fix the lattice fillings of the 
two species to $n_\uparrow=1/3$ and $n_\downarrow=2/3$.  
  
 In the extreme limit of mass imbalance, $J_{\downarrow} = 0$, (\ref{BHM}) reduces to the
 well-known Falicov--Kimball model of mobile particles in a potential created by static impurities \cite{FalicovK69}. For the 
 considered filling it can be shown via exact diagonalization that, at sufficiently low attraction $U/J_{\uparrow} \leq 2.3$, 
 the ground state realizes a crystal of \emph{trimers} formed by two $\downarrow$-bosons ``glued" together by an 
 $\uparrow$-boson in an atomic analogue of a covalent bond  (see Fig.~\ref{fig:evolution}a for a scheme of the 
 spatial arrangement). The trimer crystal is protected by a finite energy 
 gap against dislocations of the $\downarrow$-bosons, and hence it is expected to survive the presence 
 of a small hopping  $J_{\downarrow}$. We have tested this hypothesis with 
 extensive quantum Monte Carlo simulations based on the canonical
 Stochastic Series Expansion algorithm \cite{SSE, Roscilde08}. 
  Simulations have been performed on chains of size $L=30,...,120$ with periodic boundary conditions, at an inverse temperature
 $\beta J_{\downarrow}=2L/3$ ensuring that the obtained data describe the 
 zero-temperature behaviour for both atomic species.

Fig.~\ref{fig:phases}a shows the resulting ground-state phase diagram, which features
indeed an extended trimer crystal phase.   
 For $U/J_{\uparrow} \geq 2.3$, and over a large region of $J_{\downarrow}/J_{\uparrow}$ ratios,
 the ground state shows instead the progressive merger of the trimers into hexamers, dodecamers, and 
 finally into a fully collapsed phase with phase separation of the system into particle-rich and particle-free
 regions.  
  
  For $U/J_{\uparrow} \lesssim 2.3$, increasing the $J_{\downarrow}/J_{\uparrow}$ ratio allows
  to continuously tune the zero-point quantum fluctuations of the $\downarrow$-atoms 
  in the trimer crystal and 
  to increase the effective size of the trimers, whose wavefunctions start to overlap. 
  We find that, when trimers spread over a critical size of $\approx 2.8$  lattice sites, 
  they start exchanging atoms and the quantum melting of the crystal is realized.  
  The melting point is also consistent with the point at which the hopping $J_\downarrow$ 
  overcomes the energy gap to dislocations (dash-dotted line in Fig.~\ref{fig:phases}a). The resulting phase after quantum melting 
  is a one-dimensional
  superfluid for both atomic species: in this phase quasi-condensation appears,
  in the form of power-law decaying phase correlations 
   $\langle b^{\dagger}_{i,\sigma} b_{j,\sigma} \rangle \propto |r_i-r_j|^{-\alpha_{\sigma}}$, 
  which is the strongest form of off-diagonal correlations possible in interacting
  one-dimensional quantum models \cite{Giamarchibook}. Yet in the superfluid
  phase strong power-law density correlations survive,
  $\langle n_{i,\sigma} n_{j,\sigma} \rangle \propto \cos(q_{\rm tr} (r_i-r_j)) ~|r_i-r_j|^{-\beta_{\sigma}}$,
  exhibiting oscillations at the trimer-crystal wavevector $q_{\rm tr}=2\pi/3$.
  Such correlations stand as remnants of the
  solid phase, and in a narrow parameter region they even lead 
  to a \emph{divergent} peak in the density structure factor,
  $S_{\sigma}(q_{\rm tr}) \propto L^{\beta_{\sigma}}$
with $0 < \beta_{\sigma} < 1$, where 
  \begin{equation}
  S_{\sigma}(q) = \frac{1}{L} \sum_{ij} e^{iq(r_i-r_j)} \langle n_{i,\sigma} n_{j,\sigma} \rangle .
  \end{equation}
This phase, termed ``super-Tonks" 
phase in the literature on one-dimensional quantum systems \cite{sTonks},  
is a form of quasi-supersolid, in which one-dimensional superfluidity 
coexists with quasi-solid order. (Notice that true solidity corresponds to $\beta_{\sigma}=1$.)
 
 The strong competition between solid order and superfluidity in the ground-state 
 properties of this model suggests the intriguing possibility that true supersolidity might appear
 by perturbing the system out of the above equilibrium state.
 In particular we investigate the Hamiltonian evolution of the system after
 its state is prepared out of equilibrium in a perfect trimer crystal. 
 The initial state is a simple factorized state of perfect trimers (see Fig.~\ref{fig:evolution}a):
\begin{equation}
  \label{crystal}
\ket{\Psi_0} =  \bigotimes_{n=1}^{L/3} \ket{\Phi_{\rm tr}^{(3n-1)}}
\end{equation}
where the trimer wavefunction reads 
\begin{equation}
  \label{trimer}
\ket{\Phi_{\rm tr}^{(i)}} = \tfrac{1}{\sqrt{2}} b_{i \downarrow}^\dagger b_{i+1 \downarrow}^\dagger (b_{i \uparrow}^\dagger + b_{i+1 \uparrow}^\dagger) \ket{\textrm{vac}} .
\end{equation}
This state can be realized with the current technology of optical superlattices \cite{Bloch}, 
by applying a strong second standing wave component $V_{x_2} \cos^2[(k/3)x + \pi/2]$ to the 
primary wave, $V_{x_1} \cos^2(kx)$, creating the optical lattice along the $x$ direction of the chains.     
This superlattice potential has the structure of a succession of double wells separated
by an intermediate, high-energy site. Hence tunneling out of the double wells is strongly
suppressed, stabilizing the factorized state (\ref{crystal}). 
 After preparation of the system in the initial state, the second component of the 
superlattice potential is suddenly switched off ($V_{x_2}\to 0$) and the state is let
to evolve with the Hamiltonian corresponding to different parameter sets
($U/J_{\uparrow}, J_{\downarrow}/J_{\uparrow}$).
The successive time evolution over a short time interval $[0,\tau]$ with $\tau = 3 \hbar/J_{\downarrow}$ is computed using the Matrix-Product-States (MPS) algorithm 
on a one-dimensional lattice with up to $28$ sites and open boundary
conditions \cite{juanjovid}.  
 A bond dimension $D=500$ ensures that the weight of the discarded 
Hilbert space is $< 10^{-3}$.  The evolution time step $dt=5
\times 10^{-3} \hbar / J_{\uparrow}$ is chosen so as to make the
Trotter error smaller than $10^{-3}$.
We characterize the evolved state by 
averaging the most significant observables over the last portion 
of the time evolution $\tau/3$.

\begin{figure} [!h]
  \centering
  \includegraphics[width=0.9\linewidth]{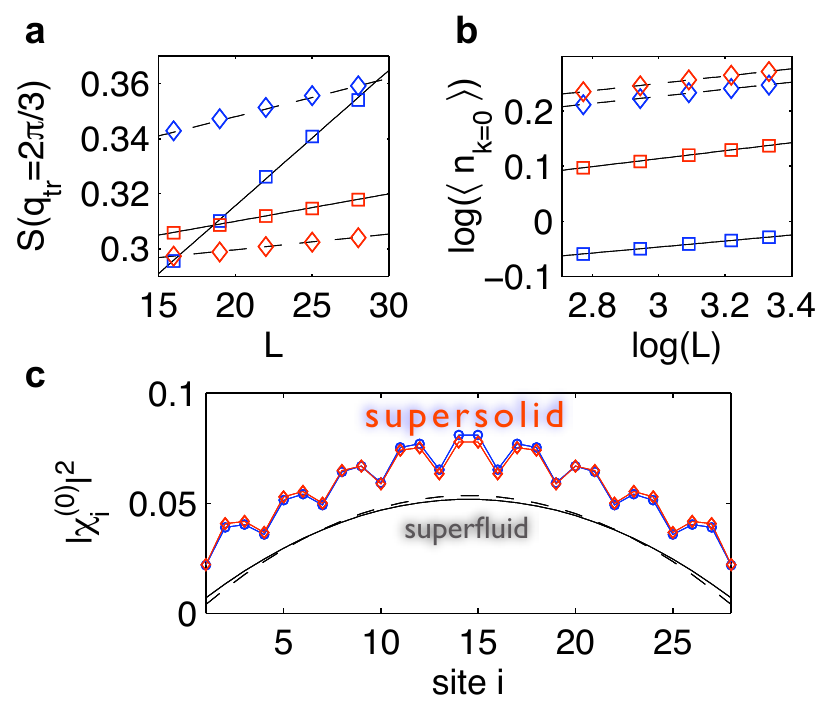}
  \vspace{-0.5cm}
  \caption{\textbf{Coexistence of solid order and quasi-condensation in the supersolid phase.} 
  \textbf{(a)} The structure factor peak $S(q_{\rm tr}=2\pi/3)$ scales linearly with system size $L$, demonstrating solid order for both bosonic species. 
  \textbf{(b)} The density peak in momentum space $\langle n^\downarrow_{k=0} \rangle$ is plotted vs. $L$ on a log-log scale, showing algebraic 
  scaling and thus quasi-condensation. 
  Boxes (diamonds) stand for particle species $\downarrow$ ($\uparrow$), respectively.
    The data represented by blue boxes in part (a) is offset by -0.2 for better visibility.
Parameters:  $J_\downarrow/J_{\uparrow} = 0.1, U/J_{\uparrow}=3.0$ (blue symbols) and 
$J_\downarrow/J_{\uparrow} = 0.15, U/J_{\uparrow}=2.5$ (red symbols).
    \textbf{(c)} Square modulus of the natural orbital $\chi_i^{(0)}$ corresponding to the largest eigenvalue of the OBDM, 
 calculated at final time $\tau$. In the supersolid regime (blue/red symbols for $\downarrow$/$\uparrow$ bosons), the natural orbital 
 shows the characteristic crystalline order. This pattern is washed out in the purely quasi-condensed regime 
 (dashed/solid curves for $\downarrow$/$\uparrow$). The supersolid data is offset by $+0.02$ for the 
 sake of visibility. Parameters:  $J_\downarrow/J_{\uparrow} = 0.1$ (supersolid), $J_\downarrow/J_{\uparrow} = 0.8$ (quasi-condensed),  
 $U/J_{\uparrow}=3.0$, $N_{\downarrow} = 18$, $N_{\uparrow} = 9$, $L=28.$ }
  \label{fig:scaling}
\end{figure}

We find three fundamentally different evolved states, whose extent
in parameter space is shown on the non-equilibrium phase diagram of Fig.~\ref{fig:phases}b:
\linebreak Firstly, a superfluid phase, in which the initial crystal structure is completely melted
by the Hamiltonian evolution, and coherence builds up in the system leading
to quasi-condensation out-of-equilibrium, namely to the appearence of 
a (sub-linearly) diverging peak in the momentum distribution
$
\langle n^{\sigma}_k \rangle = \frac{1}{L} \sum_{ij} e^{ik(r_i-r_j)} \langle b_{i,\sigma}^\dagger b_{j, \sigma}\rangle
$
 at zero quasimomentum, $\langle n^{\sigma}_{k=0} \rangle \propto L^{\alpha_\sigma}$ with $0< \alpha_\sigma < 1$. 
Despite the short time evolution, quasi-condensation of the slow $\downarrow$-bosons is probably assisted by their interaction with the faster $\uparrow$-bosons, and is observed to occur for all system sizes considered.
 \linebreak Secondly, we find a solid
 phase, in which the long-range crystalline phase of the initial state is preserved,
 as shown by the structure factor which has a linearly diverging peak at the trimer-crystal wavevector
 $S(q_{\rm tr}) \propto L$. 
 \linebreak Thirdly, an extended supersolid phase emerges, with perfect coexistence
 of the two above forms of order for both atomic species. This is demonstrated
 in Fig.~\ref{fig:scaling} a,b via the finite-size scaling of the peaks in the momentum 
 distribution and in the density structure factor. 
 In this phase, which has no
 equilibrium counterpart, the Hamiltonian evolution leads to the delocalization
 of a significant fraction of $\uparrow$- and $\downarrow$-bosons over the entire system size.
 Consequently quasi-long-range coherence builds up and the momentum distribution, which is 
 completely flat in the initial localized trimer-crystal state, acquires a pronounced peak
 at zero quasi-momentum $k=0$, as shown in Fig.~\ref{fig:evolution}b. 
 Yet the quasi-condensation order parameter $\chi_i^{(0)}$, namely the 
 natural orbital of the one-body density matrix (OBDM) $\langle b_{i,\sigma}^\dagger b_{j, \sigma}\rangle$
 corresponding to the largest eigenvalue and hosting the condensed particles,
 is spatially modulated (\emph{cf.} Fig.~\ref{fig:scaling}c), revealing the persistence of solid order
 in the quasi-condensate.  
In addition, solidity can be confirmed by direct inspection of the real-space density 
$\langle n_{i \sigma} \rangle$ (\emph{cf.} inset of Fig.~\ref{fig:evolution}b). Going from the boundaries towards 
the center, the density profiles 
of both species are modulated by the crystal structure, and the modulation amplitudes saturate at constants which 
turn out to be independent of the system size. 
 
 To gain further insight into the mechanism underlying the stabilization of a commensurate
two-species supersolid via out-of-equilibrium time evolution, we finally compare the equilibrium
 phase diagram with the non-equilibrium one.  Fig.~\ref{fig:phases}b shows that
 the superfluid/solid and superfluid/phase-separation boundaries at equilibrium 
 overlap with the threshold of formation of the supersolid out of equilibrium
 upon increasing $J_{\downarrow}/J_{\uparrow}$. This means that a quantum
 quench of the system Hamiltonian to the parameter range corresponding to a
 superfluid equilibrium ground state is a necessary condition for supersolidity to 
 dynamically set in. 
 
   \begin{figure}
  \centering
    \vspace{0.5cm}
  \includegraphics[width=1.0\linewidth]{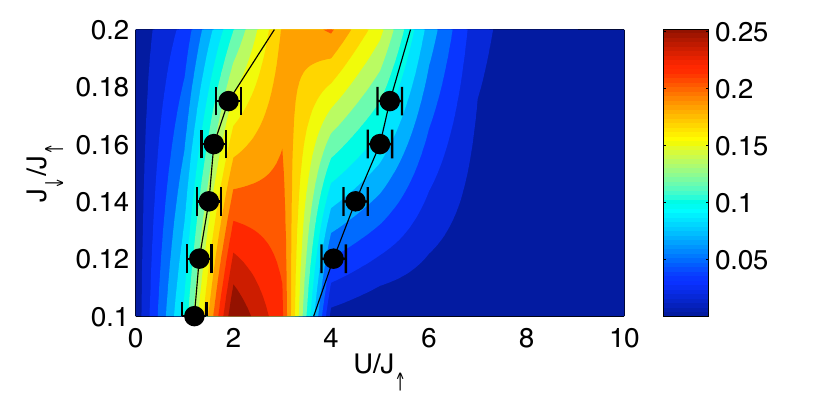}
  \caption{{\bf Overlap of the equilibrium ground state with the initial trimer-crystal state.}
 The overlap $|c_0|^2$ (contour plot) agrees well with the boundaries of the non-equilibrium supersolid phase (black symbols,
  \emph{cf.} Fig.~\ref{fig:phases}b). This suggests a superfluid ground state as a necessary condition for supersolidity to dynamically set in.
The overlap $|c_0|^2$ has been calculated via exact diagonalization on a  $L=10$ chain containing three trimers. }
  \label{fig:c0}
\end{figure}

 The key to the dynamical emergence of a quasi-condensate
 fraction in the supersolid phase is that the initial trimer-crystal state (\ref{crystal})
 has a significant overlap with the superfluid ground
 state of the final Hamiltonian after the quantum quench. 
  As shown in Fig.~\ref{fig:c0} for a small cluster with $L=10$ sites, 
  the ground-state overlap $|c_0|^2$ remains sizable
 over an extended parameter range. This is intimately connected with the 
 strong density--density correlations present in the equilibrium superfluid phase, as
 shown \emph{e.g.}\ by the appearance of a region with super-Tonks behavior.
 The excellent agreement between the region featuring supersolidity and the
 region with most pronounced overlap $|c_0|^2$ suggests the following 
 mechanism: the Hamiltonian evolution following the quantum quench dynamically
 selects the ground-state component 
 as the one giving the dominant contribution to \mbox{(quasi-)}long-range coherence.  
 In essence, while the quantum melting phase transition occurring
 at equilibrium leads to a dichotomy between solid and superfluid order, 
 the out-of-equilibrium preparation can coherently admix the excited crystalline state(s) 
 with the superfluid ground state without disrupting their respective forms
 of order (see EPAPS).  It is tempting to think that a similar preparation scheme of supersolid
 states can work in other systems displaying solid--superfluid phase
 boundaries at equilibrium. 
 
The supersolid transient state is an exquisitely non-equilibrium state, because no order can
survive at finite temperature in 1D systems with short-range interactions.  
An intriguing question arises then: in the long-time limit  $\tau \gg 3 \hbar/J_{\downarrow}$
(which is only accessible numerically on very small system sizes) does supersolidity survive 
or is long-range order ultimately destroyed by thermalization?
Recent numerical studies point towards the failure of other strongly correlated one-dimensional quantum systems to thermalize \cite{Kollathman}.  
We have considered the asymptotic time limit using exact diagonalization for a small system (see EPAPS).
 These exact results suggest that supersolidity persists and the system does not converge
 to an equilibrium thermal state (in fact even thermalization in the microcanonical ensemble, proposed in Ref.~\cite{Rigoletal08}, 
 does not seem to occur in our system). 
Whether the absence of thermalization survives when taking the thermodynamic limit remains an open question, whose
answer at the moment can only rely on experiments. 
  
 The observation of the supersolid state prepared via the dynamical scheme proposed in 
 this Letter is directly accessible to several setups in current optical-lattice experiments. 
 The fundamental requirement to explore the phase diagrams of our model, Fig.~\ref{fig:phases}, 
 is the existence of a stable bosonic mixture with mass imbalance and  
 interspecies interactions that can be tuned to the attractive regime via
 a Feshbach resonance.
 This requirement is met in spin mixtures of, \emph{e.g.}, $^{87}$Rb atoms
 in different hyperfine states, 
 which acquire a spin-dependent effective mass when loaded in an optical 
 lattice \cite{Mandel}, and for which Feshbach resonances have been
 extensively investigated \cite{Marteetal02}. Moreover recently discovered
 Feshbach resonances in ultracold heteronuclear bosonic mixtures ($^{87}$Rb-$^{133}$Cs, $^7$Li-$^{87}$Rb, 
  $^{41}$K-$^{87}$Rb,  $^{39}$K-$^{87}$Rb etc. \cite{Chinetal08}, 
 the latter recently loaded in optical lattices \cite{Catanietal08})
enlarge even further the number of candidate systems to implement the Hamiltonian (\ref{BHM}). The hardcore-repulsive regime can be easily
accessed in deep optical lattices \cite{Paredes04}. After
preparation of the trimer crystal via an optical superlattice \cite{Bloch},
the onset of coherence
in the supersolid state, attained after a short hold time corresponding to $\approx $ 2-3 
hopping events of the slower particles  ($\approx $ 1-10 ms), can be monitored by 
time-of-flight measurements of the momentum distribution.
The rapid onset of coherence allows the experimental detection of supersolidity well 
before decoherence effects become important. 
On the other hand, the persistence of the crystalline structure can be probed by  
resonant Bragg scattering \cite{Bragg}. While experimentally the initial
state will be always a mixed one and not the pure state in (\ref{crystal}),
we observe that mixedness of the initial state does not
disrupt supersolidity in the evolved state.

We thank J. J. Garcia-Ripoll, M. Roncaglia, and R. Schmied for helpful discussions.
This work is supported by the European Union through the SCALA integrated
project.

\newpage

\begin{appendix}
\section{Supplementary Material}

In the following we present exact calculations for a small system which elucidate the special nature of the initial trimer crystal state after the quench, superimposing the superfluid (and quasi-condensed) ground state with selected crystalline excited states.
Furthermore, we compare the results for the asymptotic state of the time evolution with 
thermal states in both the canonical and microcanonical ensembles. Our results indicate 
that thermalization may not occur in our system.

\subsection{Time evolution of the initial trimer-crystal state}
 We discuss here in more detail the time evolution of the initial
 trimer-crystal state into a supersolid state.   
 The initial state, equation (3) of the main Letter, can be decomposed into the eigenstates of the final Hamiltonian ${\cal H} |E_a\rangle = E_a |E_a \rangle$ as:
\begin{equation}
|\Psi(t=0)\rangle=  \sum_{a} c_a |E_a\rangle . 
\end{equation}
 The time-evolved state is then: 
\begin{equation}
|\Psi(t)\rangle=  \sum_{a} c_a e^{-i\omega_{a} t} |E_a\rangle
\end{equation}
where $\omega_a = E_a / \hbar$. 

The expectation value of any operator $A$ can be thus written as
\begin{eqnarray}
\langle A \rangle_{t} &=&   \sum_{a} |c_a|^2 \langle E_a|  A |E_a \rangle    \\
&+& \sum_{a\neq b} 2~{\rm Re}\left[ \langle  E_a| A |E_b \rangle c_a^* c_b e^{i(\omega_a-\omega_b)t} \right]  \nonumber
\end{eqnarray}
approaching the ``diagonal ensemble" [18] or steady state for a large time $t \rightarrow \infty$,
\begin{equation}
\langle A \rangle_{\infty} =   \sum_{a} |c_a|^2 \langle E_a|  A |E_a \rangle.
\label{diagonal}
\end{equation}

 We now specify the discussion to the case in which the system is evolved with a quantum Hamiltonian
whose ground state is both a superfluid and a quasi-condensate. 
If the initial trimer-crystal state has a significant overlap with the quasi-condensed ground state, namely if $c_0$
is not negligible, then one can expect that the phase correlator of the steady state, 
corresponding to $A \equiv b_{i,\sigma}^\dagger b_{j,\sigma}$, will be dominated by the ground-state
contribution, so that (quasi-)long-range order sets in. At the same time, 
the initial state has by construction a significant projection on excited states  $|E_{a>0}\rangle$ 
with long-range crystalline correlations, provided that these states exist in the
Hamiltonian spectrum.
Under this assumption, the  density-density correlator, corresponding to $A \equiv n_{i,\sigma} n_{j,\sigma}$,
will remain long-ranged in the steady state; this fact, combined with \mbox{(quasi-)}long-range  
phase coherence, gives rise to supersolidity.

  \begin{figure}[h!]
  \centering
  \includegraphics[width=0.9\linewidth]{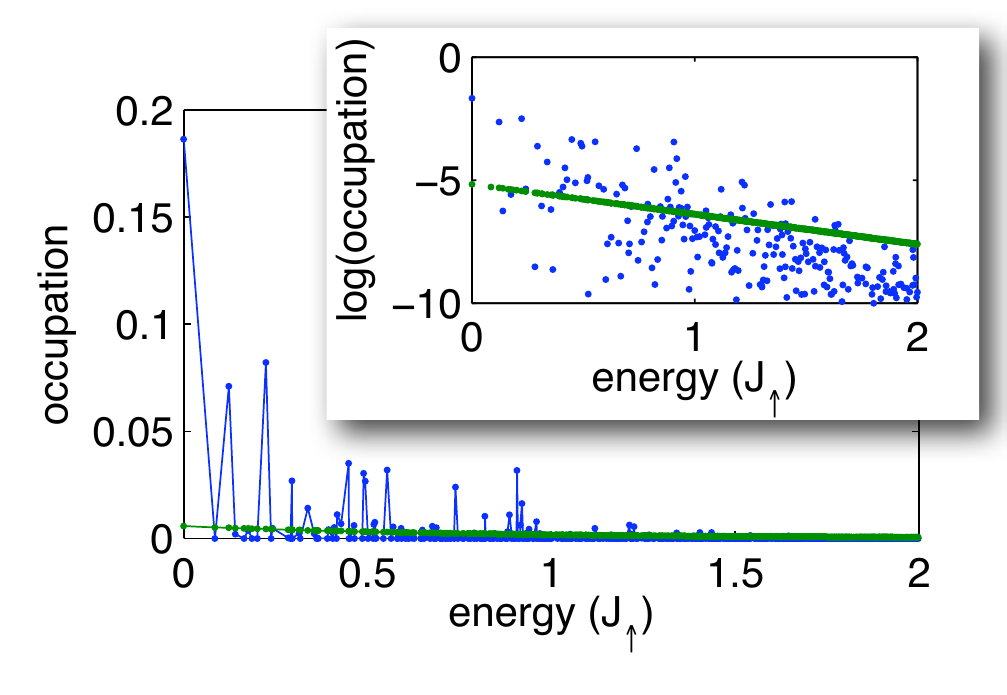}
  \caption{\textbf{Diagonal vs. thermal probability distributions.}
  The occupations of the diagonal ($|c_a|^2$ in blue) and canonical ($|d_a|^2$ in green) ensembles are plotted as a function of the eigenstate energies (offset from $E_{GS}$).
  Contrary to the thermal, continuous distribution, the trimer-crystal state emphasizes certain eigenstates, while it suppresses others.
  The (superfluid) ground state contribution present in the trimer-crystal state is enhanced by a factor of $\approx 20$ compared with the thermal contribution.
 Most of the amplified excited states indeed show a crystalline structure with the correct periodicity, or contain density peaks at the right positions to build up the final crystal.
  Inset: The same distributions on a log-lin scale. The deviation of the diagonal from the thermal ensemble is even better visualized here.  }
  \label{epaps1}
\end{figure}

Making use of exact diagonalization on a $L=10$ chain with open boundary conditions,
we have systematically investigated the overlap $c_0$ between the perfect trimer-crystal
state and the Hamiltonian ground state for different points in parameter space. The results are shown in Fig. 4 of the main Letter, and compared
with the phase boundaries of the non-equilibrium phase diagram, Fig. 2b of the main Letter.
We observe that the non-equilibrium supersolid phase is in striking correspondence
with the parameter region where $c_0$ is largest, suggesting that the 
above analysis of the onset of supersolidity is quantitatively correct. Note that the time evolution discussed in the main Letter is restricted to finite times, while we focus here on the asymptotic case $t \rightarrow \infty$.

\subsection{Comparison of the asymptotic state with thermal states}

The diagonal-ensemble expectation value of equation (\ref{diagonal}) is here compared with a thermal average in the canonical ensemble
\begin{equation}
\langle A \rangle_{T} =   \sum_{a} |d_a|^2 \langle E_a|  A |E_a \rangle,
\end{equation}
with $|d_a|^2=\exp(-E_a/k_B T) / Z$ the Boltzmann weights, $k_B$ Boltzmann's constant, $T$ the temperature and $Z=\sum_a \exp(-E_a/k_B T)$ the normalizing partition function.
In addition, we introduce for comparison the statistical average in the microcanonical ensemble
\begin{equation}
\langle A \rangle_{E_{\rm in},dE} =   \sum_{E_{\rm in}-dE<E_a<E_{\rm in}+dE} 1/N_m  \langle E_a|  A |E_a \rangle,
\end{equation}
which averages over eigenstates within an energy window $\pm dE$ around the inital energy $E_{\rm in} = \langle\Psi(t=0)| {\cal H} |\Psi(t=0)\rangle$. 
$N_m$ is the number of eigenstates contained in that energy window.
 
In order to compare the diagonal with the canonical and microcanonical ensembles, we have chosen to exactly diagonalize a system of three trimers ($N_\downarrow=6, N_\uparrow=3$) in an open chain of $L=10$ sites. We present in the following results for the parameter pair ($J_\downarrow=0.2 J_\uparrow, U=3 J_\uparrow$), where supersolidity exists according to our non-equilibrium phase diagram, Fig. 2b of the main Letter. 
Under these conditions, the ground state energy yields $E_{GS} \simeq -13.2 J_\uparrow$, while the initial trimer-crystal state carries an energy $E_{\rm in} = -12 J_\uparrow$. In order to determine the correct temperature for the canonical ensemble, we have varied $T$ until the condition $ \langle {\cal H} \rangle_{T}=E_{\rm in}$ was met. This analysis yielded $k_B T \simeq 0.82 J_\uparrow$, which we use henceforth for the comparison with the canonical averages.

 \begin{figure}[h]
  \centering
  \includegraphics[width=0.8\linewidth]{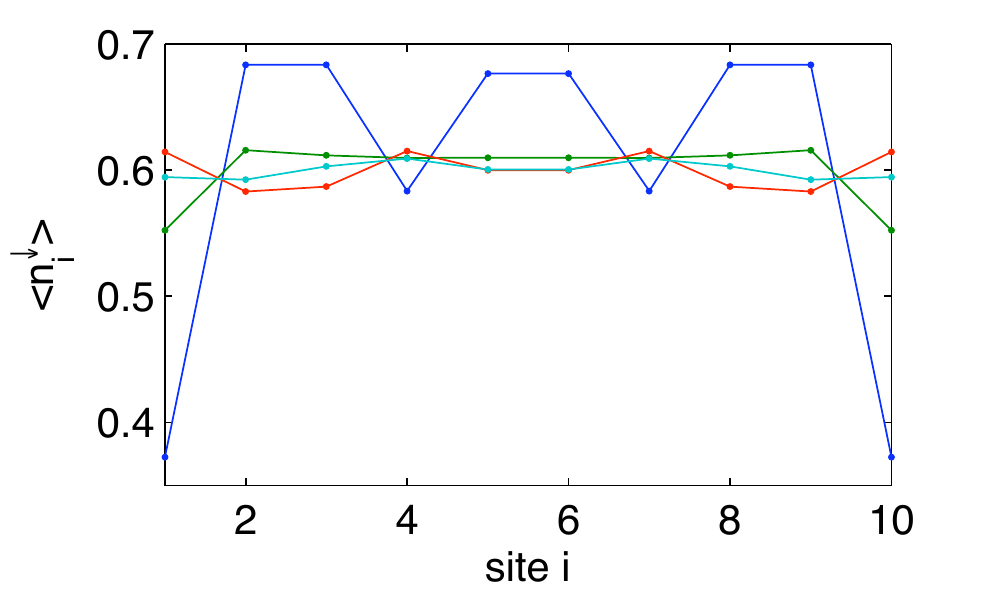}
  \caption{\textbf{Real-space density $\langle n_i^\downarrow\rangle$ in all three ensembles.} While the diagonal ensemble $\langle n_i^\downarrow\rangle_\infty$ (blue) shows a clear crystalline pattern, this structure is washed out completely in the canonical ensemble $\langle n_i^\downarrow\rangle_{T=0.82 J_\uparrow / k_B}$ (green). Results for the microcanonical ensemble $\langle n_i^\downarrow \rangle_{E_{\rm in},dE}$ are shown for energy windows $dE=0.2 J_\uparrow$ (red) and  $dE=0.6 J_\uparrow$ (cyan). All thermal ensembles deviate strongly from the density structure at time $t \rightarrow \infty$ (diagonal ensemble).}
    \label{epaps2}
\end{figure}

 \begin{figure}[h]
  \centering
  \includegraphics[width=0.8\linewidth]{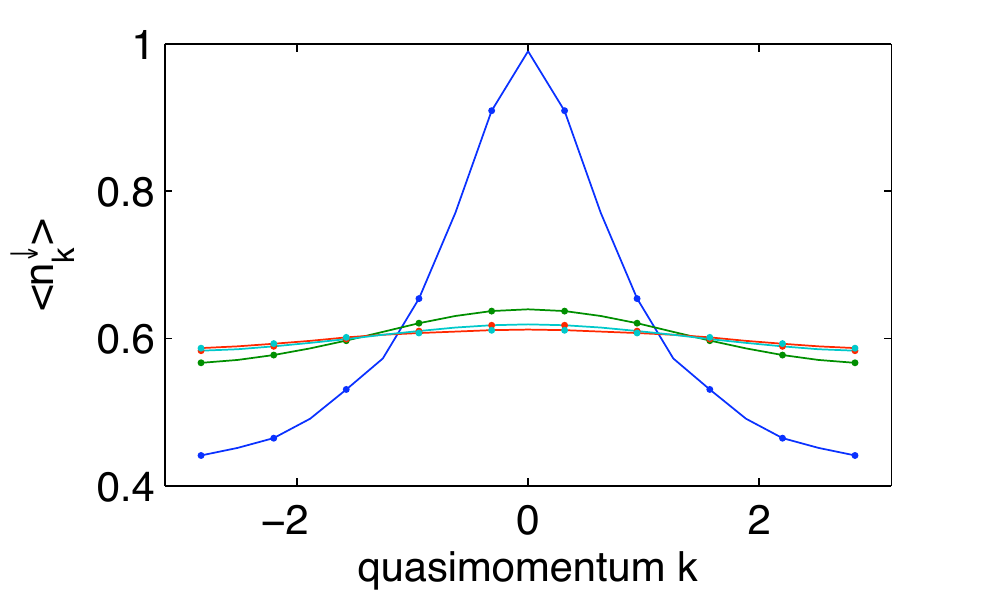}
  \caption{\textbf{Momentum profile $\langle n_k^\downarrow\rangle$ in all three ensembles.} Due to the significant weight attributed
  to the ground state, the diagonal ensemble $\langle n_k^\downarrow\rangle_\infty$ (blue) features an enhanced quasi-condensation peak at $k=0$. This peak is suppressed in all thermal ensembles $\langle n_k^\downarrow\rangle_{T=0.82 J_\uparrow / k_B}$ and $\langle n_k^\downarrow \rangle_{E_0,dE}$ (same colouring scheme as in Fig. \ref{epaps2}).  }
    \label{epaps3}
\end{figure}

Fig. \ref{epaps1} compares the diagonal ensemble induced by the initial trimer-crystal state with a thermal, canonical ensemble. The trimer-crystal state has a finite projection on the quasi-condensed ground state as well as on distinct excited states. Further inspection into those excited states shows that their characteristic density profiles matches the crystal structure of the initial state. Hence
the selection of excited states in the diagonal ensemble is fundamentally governed by the broken translational invariance present in the initial state. In contrast, the canonical ensemble averages over all eigenstates regardless of their displaying crystalline order, a fact which makes the loss of the crystalline structure unavoidable.

The density profiles (for the $\downarrow$-bosons) shown in Fig. \ref{epaps2} corroborate the previous statements. The diagonal ensemble induced by the trimer crystal is compared here with thermal averages in both the canonical and microcanonical ensembles. While the density profile in the diagonal ensemble still displays the ``memory effect" of the initial crystalline state, the thermal states exhibit only small density modulations (in the microcanonical ensemble) or no modulation at all (in the canonical ensemble). Furthermore, the momentum profiles shown in Fig. \ref{epaps3} underline a non-thermalization scheme of the time-evolved crystal state. While the density profile of the diagonal ensemble exhibits a pronounced peak at quasimomentum $k=0$, this peak is almost completely washed out for the thermal ensembles.

In view of the two observables discussed here, a thermalization of the evolved trimer crystal state can be excluded, at least for the finite-size system we are considering. 
This confirms the observations of  ``non-thermalization" in other 1D finite-size systems [17].

Our exact diagonalization study is limited to a small cluster, and it cannot exclude a priori that thermalization 
appears for larger system sizes: this would require that the diagonal ensemble converges to the microcanonical
one, which ultimately converges to the canonical ensemble in the thermodynamic limit.

\subsection{Numerical results of long-time evolutions}

\begin{figure}
  \centering
  \includegraphics[width=0.8\linewidth]{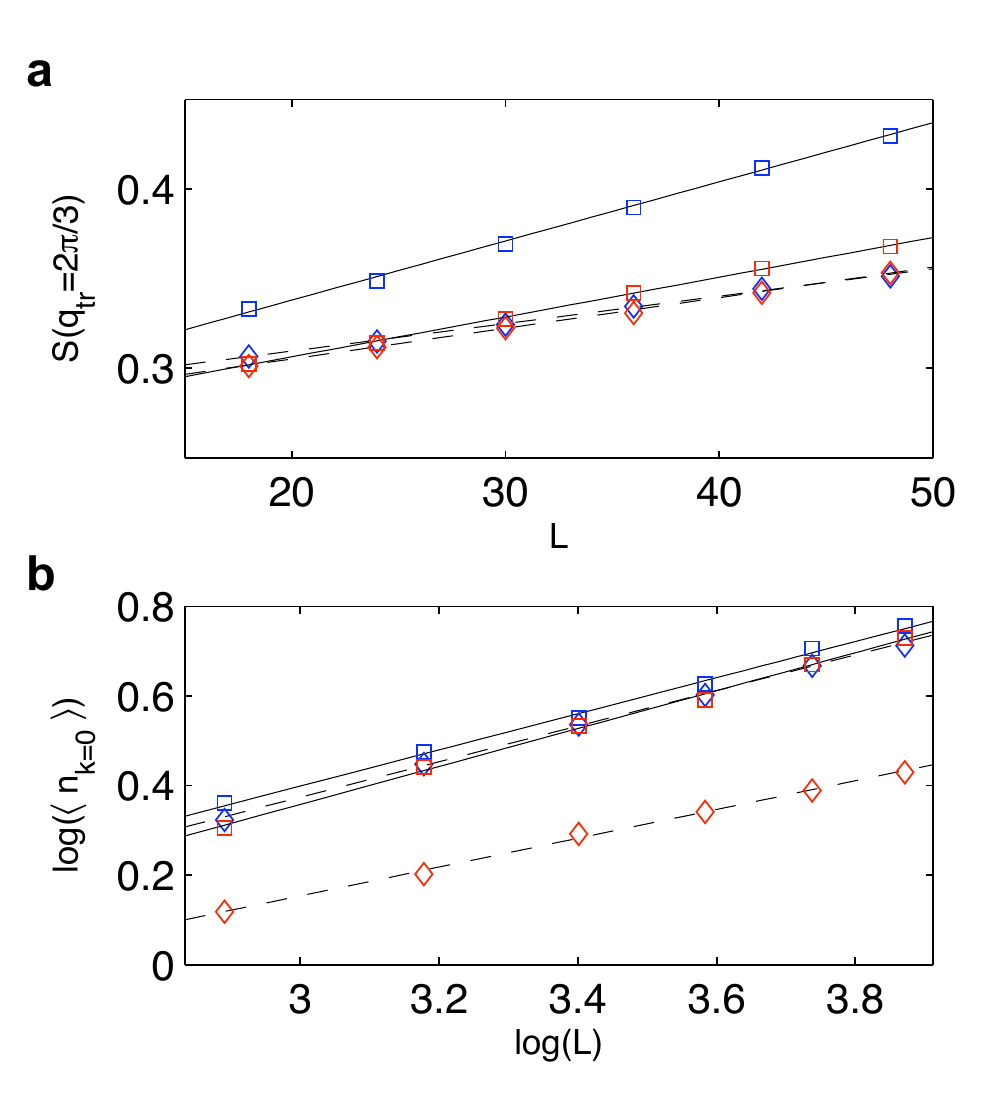}
  \vspace{-0.5cm}
  \caption{\textbf{Scaling analysis of the long-time evolution data.} 
  \textbf{(a)} Structure factor peak $S(q_{\rm tr}=2\pi/3)$; 
  \textbf{(b)} Quasi-condensate peak $\langle n_{k=0} \rangle$. 
  Boxes (diamonds) stand for particle species $\downarrow$ ($\uparrow$), respectively.
 Parameters:  $J_\downarrow/J_{\uparrow} = 0.15, U/J_{\uparrow}=2.5$ (blue symbols) and 
 $J_\downarrow/J_{\uparrow} = 0.40, U/J_{\uparrow}=9.0$ (red symbols).}
  \label{epaps4}
\end{figure}

 Here we present an example of the scaling analysis for the results of a long-time
 evolution up to  $\tau = 150 \hbar/J_{\downarrow}$. Fig. \ref{epaps4} shows
 that observables averaged over the last $\tau/4$ interval of the time evolution 
 display the characteristic one-dimensional supersolid scaling, analogous to --
 but much more marked than -- the one observed at short times 
 (compare Fig. 3a-b of the main Letter). Indeed we observe a linear scaling
 of the structure factor peak  $S(q_{\rm tr}=2\pi/3)$ with system size, 
 typical of solid order, 
 and an algebraic sub-linear scaling of the condensed atoms, signaling
 quasi-condensation. Repeating
 this scaling analysis for a fine mesh of parameter space leads to the
 confirmation of the supersolid phase shown in Fig. 2b of the main Letter. 

  A word of caution is necessary in the case of long-time evolutions. 
  The truncation of the Hilbert space, inherent in all numerical
 algorithms for time evolution not based on full exact diagonalization [16], 
 has the general effect that the accuracy 
 of the results rapidly degrades with time, and the instantaneous measurements 
 become practically unreliable in the long-time limit. 
 We observe, however, that observables averaged over time intervals $> \tau/10$ do converge with high precision upon variation of the bond dimension $D$.
 These time averaged results
 are indeed the object of the scaling analysis in Fig. \ref{epaps4}. 
 
\end{appendix}

\end{document}